\documentstyle[12pt]{article}
\input epsf      
\epsfverbosetrue

\newcommand \preprint {1}

\newcommand \s {\sigma}
\newcommand \p {\prime}
\newcommand \no {\noindent}
\newcommand \qed{\vrule height5pt width5pt}
\newcommand \boundary {\partial}

\newcommand \figa {1 \ } 
\newcommand \figb {2 \ } 
\newcommand \figc {3 \ } 
\newcommand \figd {4 \ } 
\newcommand \fige {5 \ } 
\newcommand \figz {6 \ }
\newcommand \figf {7 \ } 
\newcommand \figg {8 \ } 

\begin{document}

\title{Majority Rule at Low Temperatures on the 
Square and Triangular Lattices }

\author{Tom Kennedy
\\Department of Mathematics
\\University of Arizona
\\Tucson, AZ 85721
\\ email: tgk@math.arizona.edu
}
\maketitle

\begin{abstract}
We consider the majority rule renormalization group transformation 
applied to nearest neighbor Ising models. For the square lattice 
with 2 by 2 blocks we prove that if the temperature is sufficiently 
low, then the transformation is not defined. 
We use the methods of \cite{efs}, who proved the renormalized measure is 
not Gibbsian for 7 by 7 blocks if the temperature is too low. 
For the triangular lattice we prove that a zero temperature 
majority rule transformation may be defined. The resulting renormalized 
Hamiltonian is local with 14 different types of interactions.
\end{abstract}

\vskip 0.7 in

{\leftskip=30 pt \rightskip= 30 pt \noindent
{\bf Keywords:} 
majority rule renormalization group transformation,
non-Gibbsian measures.
\par}

\vskip 0.5 in

{\leftskip=30 pt \rightskip= 30 pt \noindent
{\bf Archive:}  cond-mat/9605104
\par}

\newpage

\section{Introduction} \setcounter{equation}{0}

The majority rule renormalization group transformation is formally 
defined by 
\begin{equation}
e^{- H^\p(\s^\p)} = \sum_\s T(\s,\s^\p) \, e^{-H(\s)} \label{eqa}
\end{equation}
Here $\s$ are the spins in the original system and $\s^\prime$ are the 
block spins (or image spins). All spins are Ising spins, i.e., take on the
values $-1,+1$. $H$ is $\beta$ times the original Hamilton which 
throughout this paper will just be the ferromagnetic nearest neighbor 
Hamiltonian. $T(\s,\s^\p)$ is the kernel for the majority rule transformation. 
If the lattice and block size are such that the number of spins in a 
block is odd, then this kernel only takes on the values 0 and 1.
The kernel is 1 if in every block a majority of the spins agree with 
the block spin, and 0 otherwise. If the number of spins in a block is 
even, then there can be a ``tie'' in a block. In such systems 
the kernel is zero  if there is a block with a clear majority and 
the majority disagrees with the block spin. If every block has either 
a clear majority which agrees with the block spin or a tie, then the 
kernel is equal to $2^{-n}$ where $n$ is the number of ties. 
These factors of ${1 \over 2}$ are included so that $T(\s,\s^\p)$ is a 
probability kernel, i.e., for every $\s$,
$\sum_{\s^\p} T(\s,\s^\p) = 1$ .
Eq. (\ref{eqa}) formally defines a new Hamiltonian $H^\p$ for the block spins. 

Eq. (\ref{eqa}) cannot be used to directly define $H^\p(\s^\p)$ in the 
infinite volume limit. One must use (\ref{eqa}) for finite volumes and then
try to take the infinite volume limit. Another approach, emphasized in 
\cite{efs}, is to apply the majority rule transformation to infinite
volume Gibbs measures. The renormalized measure is always defined, but
one must now deal with two problems. 
First, the original Hamiltonian may have more
than one infinite volume Gibbs measure. Second, the renormalized measure 
may not be the Gibbs measure of any reasonable Hamiltonian. 
Van Enter, Fernandez and Sokal \cite{efsa,efsb,efs} 
used ideas of Griffiths and Pearce
\cite{gpa,gpb,g} and Israel \cite{i} to prove that at low temperatures
the renormalized measure is non-Gibbsian for majority rule for a variety of
block sizes, the smallest being 7 by 7. 
The idea behind the proof is to find a special block spin configuration such 
that when the system of original spins is conditioned on this block spin
configuration, there is a phase transition.
Loosely speaking, the strong correlations in the 
constrained original spin system then prevent the renormalized measure from 
being quasilocal. 

There now exists a large collection of examples
in which the renormalized measure is not Gibbsian. A review and extensive
bibliography may be found in \cite{ve}. In many of these examples, including
the case of 2 by 2 majority rule considered here, the trouble is caused 
by block spin configurations (like the checkerboard one) that will never
be seen at low temperature. 
Insisting that the renormalized measure be quasilocal uniformly in the block
spin configuration may be asking too much. By using a weaker definition of the 
renormalized Hamiltonian it is often possible to prove a renormalized
Hamiltonian may be defined in cases where the renormalized measure
is not uniformly quasilocal \cite{d,mv}.

To show that the transformation is not defined for 2 by 2 blocks,
we follow the method of \cite{efs} closely. 
The special block spin configuration that they use for 
7 by 7 blocks is the ``doubly alternating'' configuration.
This configuration consists of 2 by 2 groups of block spins of the same
sign which alternate, i.e., for each such group of four block spins 
the four adjacent groups of four block spins have the opposite sign.
With this block spin constraint the original spins have two ground states.
In one ground state most of the spins are $+1$, but there are 10 by 10
islands of $-1$ arranged so that the majority rule constraint is satisfied.
The special block spin 
configuration we use is the alternating or checkboard configuration in which
every pair of nearest neighbor block spins are not equal.
For 2 by 2 blocks with this block spin configuration we will show that
the original spins have four ground
states - the four ``strip'' states. One of them is shown in figure \figa. 
Given the methods of \cite{efs},
the only nontrivial part of our proof is to show that these are indeed 
the ground states and a ``Peierls condition'' is satisfied. 
We do this by showing that the Hamiltonian with the majority rule
constraint can be written as an ``$m$-potential'' \cite{hs}.
To verify that our rewritten form of the constrained Hamiltonian
is indeed an $m$-potential we enlist the help of a computer. 
Our final result is theorem 4.5 of \cite{efs} with ``7 x 7'' 
replaced by ``2 x 2''. For the convenience of the reader we restate 
the theorem. In the following, $\mu T$ denotes the probability measure
on the block spins which is obtained from the Gibbs measure $\mu$ 
for the originial spins and the transformation $T$ in the usual way
\cite{efs}.

\medskip

\no {\bf Theorem 1:} 
(For 7 by 7 blocks this is theorem 4.5 of \cite{efs}.)
For all $\beta$ sufficiently large, the following holds: Let $\mu$ 
be any Gibbs measure for the two-dimensional Ising model with 
nearest neighbor coupling $\beta$  and zero magnetic field. Let $T$ be the 
majority-rule transformation on 2 x 2 square blocks. Then the measure
$\mu T$ is not consistent with any quasilocal specification. 
In particular, it is not the Gibbs measure of any uniformly convergent
interaction.

\bigskip

Next we turn to our second result.
In a finite volume we can take the zero temperature limit of (\ref{eqa})
to obtain 
\begin{equation}
H^\p(\s^\p) = \min_{\s:T(\s,\s^\p) \ne 0} H(\s) \label{eqb}
\end{equation}
(In (\ref{eqa}) the inverse temperature $\beta$ is hidden in $H$, so
in taking this limit we must divide $H^\prime$ by $\beta$.)
We can then ask if this zero temperature majority rule transformation
has an infinite volume limit, i.e., if $H^\p(\s^\p)$ has an infinite 
volume limit which belongs to some reasonable Banach space of Hamiltonians.
If one looks at the argument which shows that the majority rule 
transformation is not defined at low temperature for 2 by 2 and 7 by 7
blocks, it is easy to adapt it to show that (\ref{eqb}) does not have 
a nice infinite volume limit in this case. 
For the triangular lattice the situation is quite different. We
will prove that not only does (\ref{eqb}) have an infinite volume limit, but
the renormalized Hamiltonian is a {\t local} function of the block 
spins. The precise result is as follows.
A Hamiltonian is said to be local if it contains 
only a finite number of terms up to translations.
In the following we work with finite volumes which are unions of 
blocks and which admit periodic boundary conditions. 
This last condition means that the finite volume and translations of 
it tesselate the lattice. 

\medskip

\no {\bf Theorem 2:} For finite volumes $\Lambda$ which 
admit periodic boundary conditions, 
define $H_\Lambda^\p(\s^\p)$ by (\ref{eqb}) where $H(\s)$ is the nearest
neighbor ferromagnetic Hamiltonian for the original spins in $\Lambda$ 
with periodic boundary conditions.
There is a local translation invariant Hamiltonian $H^\p(\s^\p)$ on 
the block spins such that for sufficiently large volumes $\Lambda$,
$H_\Lambda^\p(\s^\p)$ equals the restriction of $H^\p(\s^\p)$  to 
$\Lambda$ with periodic boundary conditions.
(The local renormalized Hamiltonian is given in table 1.)

\bigskip

Of course, this theorem does not prove anything about majority rule
on the triangular lattice for low but nonzero temperatures. 
However, it does show that the argument used to prove the
transformation is not defined for 7 by 7 and 2 by 2 blocks on the square 
lattice will not work for the triangular lattice. The theorem suggests
the possibility that majority rule is actually defined for 
the triangular lattice at low temperatures, possibly for all temperatures.

\newpage

\no {\bf 2. Square lattice with 2 by 2 blocks}

\bigskip

Consider the checkerboard block spin configuration (\cite{efs} calls
this configuration the fully alternating configuration.)
We will show that with the constraint imposed by this block spin 
configuration, the system of original spins has four 
periodic ground states. One of them is shown in 
figure \figa. The other three are obtained by rotating this one by 90 degrees
and by applying a global spin flip to these two spin configurations.
We will refer to these four states as the {\it strip states}.

\ifodd \preprint 
{
\epsfbox[-100 -10 200 250]{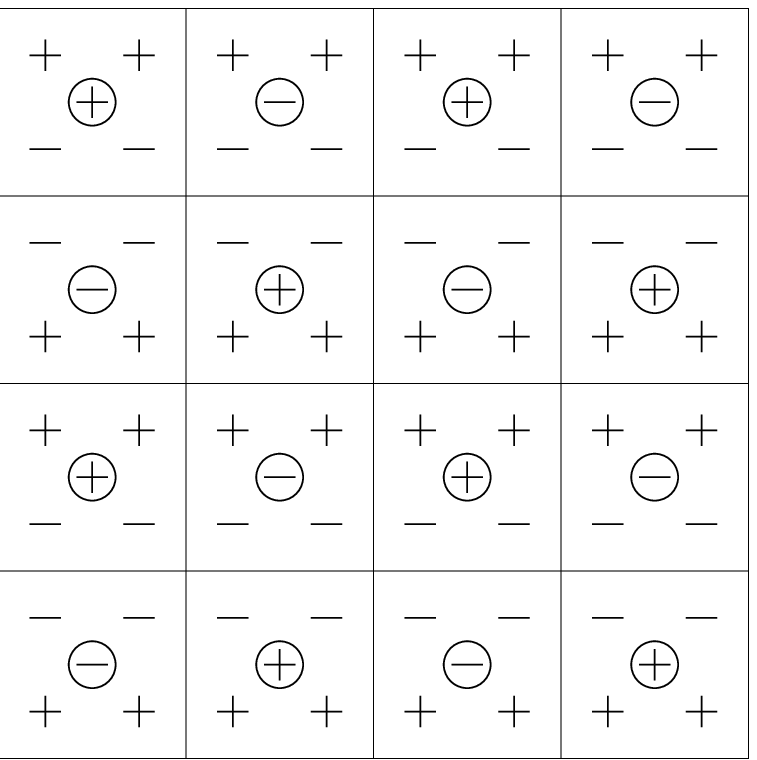}

{\leftskip=20 pt \rightskip= 20 pt \noindent
\no Figure \figa : Circled spins are block spins, uncircled spins are original
spins. The block spins are in the checkerboard configuration, and the 
original spins are in one of the four ground states which we call
``strip states''.
\par}

\bigskip
\bigskip
}
\else { } \fi

A Hamiltonian $H$ may be written in many ways as 
\begin{equation}
H= \sum_A \Phi_A
\end{equation}
where $A$ is summed over finite subsets up to some fixed size and 
$\Phi_A$ is a function of the spins in $A$. Such a decomposition is
said to be an $m$-potential if there is a configuration $\s$ such that
for every $A$
\begin{equation}
\Phi_A(\s) = \min \Phi_A
\end{equation}
In other words one can find a single configuration which simultaneously
minimizes every term in the decomposition of the Hamiltonian.

\medskip

\no {\bf Proposition 3:} $H$ may be written as an $m$-potential. Furthermore,
the only configurations which simultaneously minimize every term 
in this representation of the Hamiltonian are the four strip states.

\medskip

\no {\bf Proof:} We will take the original Hamiltonian to be 
\begin{equation}
H=\sum_{<ij>} (1-\sigma_i \sigma_j)/2 \label{eqz} 
\end{equation}
so that a pair of nearest neighbor spins that agree has energy 0 and
a pair that disagrees has energy 1. Our  representation of this
Hamiltonian that shows that it is an $m$-potential is rather 
complicated, so we will motivate it by showing why a natural simpler
representation is not an $m$-potential. 
Divide the lattice into 4 by 4 squares 
so that each square contains 4 of the 2 by 2 blocks 
used by the majority rule. 

\ifodd \preprint 
{
\epsfbox[-100 -10 200 250]{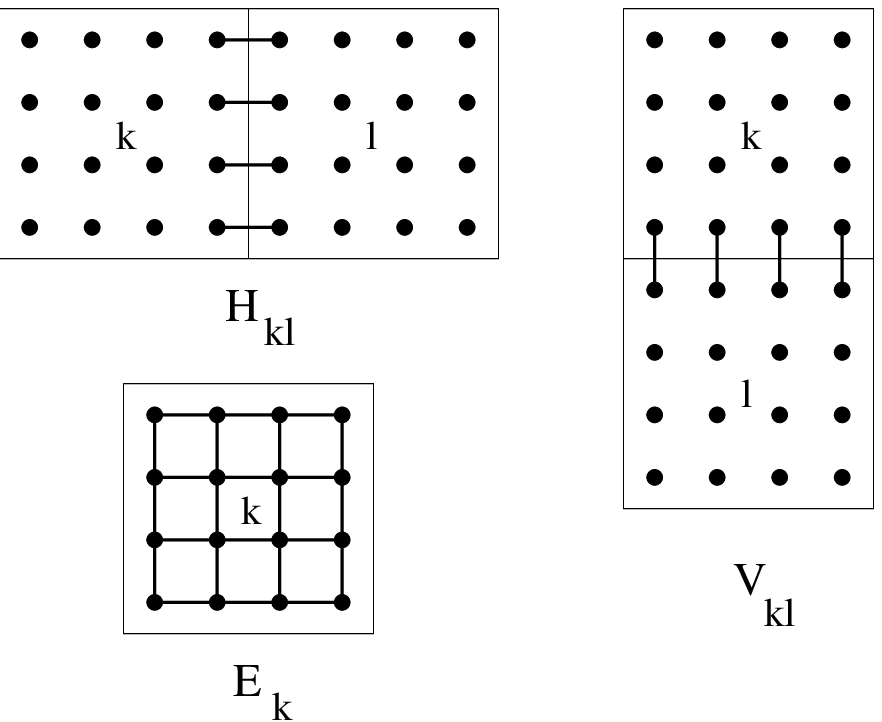}

{\leftskip=20 pt \rightskip= 20 pt \noindent
\no Figure \figb : The original Hamiltonian is the sum of the three
types of terms $E_k$, $H_{kl}$, and $V_{kl}$. The block spins are not shown
here.
\par}

\bigskip
\bigskip
}
\else { } \fi

\no Let $E_k$ be the sum of the terms in (\ref{eqz}) for
which the bond $<ij>$ is entirely in square $k$. For squares $i$ and $j$ 
which are ``horizontally adjacent'', i.e., they share a vertical edge,
let $H_{kl}$ be the sum of the terms such that one endpoint of the bond
$<ij>$ is in square $k$ and the other in square $l$. For ``vertically 
adjacent'' squares $k$ and $l$, $V_{kl}$ is defined similarly.
Figure \figb shows $E_k$, $H_{kl}$ and $V_{kl}$. 
With these definitions, 
\begin{equation}
H= \sum_k E_k +  \sum_{<kl>:hor} H_{kl} +  \sum_{<kl>:ver} V_{kl}
\label{eqaa} 
\end{equation}
The first sum is over squares $k$. The second is over horizontally
adjacent squares $k$ and $l$, and the third is over vertically adjacent
squares $k$ and $l$. Each pair of adjacent squares appears only once
in the above. We make the convention that in $H_{kl}$, $k$ is the left
square and $l$ is the right square. In $V_{kl}$, $k$ is the upper square 
and $l$ is the lower square. 
In a strip state, $H_{kl}$ and $V_{kl}$ are always
zero, and $E_k=8$. Figure \figc gives a configuration which 
gives a lower value for $E_k$, and thus this decomposition
of the Hamiltonian is not an m-potential.

\ifodd \preprint 
{
\epsfbox[-100 -10 200 150]{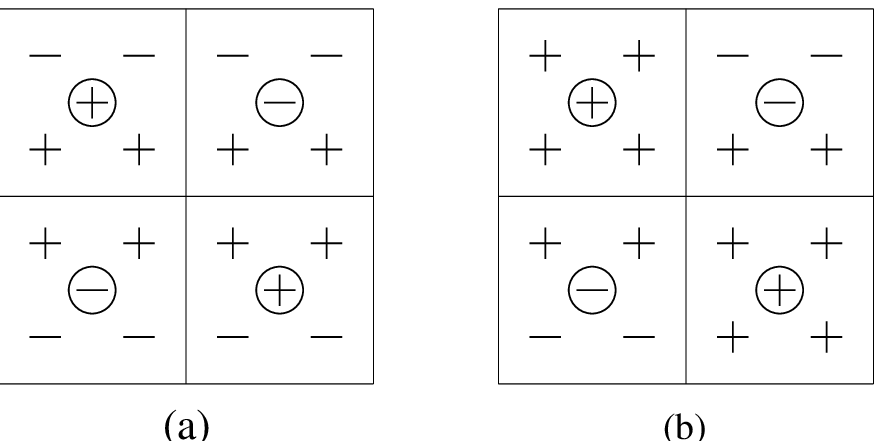}

{\leftskip=20 pt \rightskip= 20 pt \noindent
\no Figure \figc : Example showing that the strip states do not 
give the minimum of $E_k$. 
The configuration in (b) has lower $E_k$ than the strip state shown
in (a). 
\par}

\bigskip
\bigskip
}
\else { } \fi

To modify the above decomposition of the Hamiltonian to give an $m$-potential,
we introduce four functions $L_k,R_k,U_k,D_k$, each of which only
depends on the spins in square $k$. ($L,R,U,D$ stand for
left, right, up and down.) Define
\begin{eqnarray*}
\hat E_k &&=E_k+L_k+R_k+D_k+U_k \\
           \hat H_{kl} &&=H_{kl}-R_k-L_l \\
           \hat V_{kl}&&=V_{kl}-D_k-U_l \\
\end{eqnarray*}
Then we have 
\begin{equation}
H= \sum_k \hat E_k + \sum_{<kl>:hor} \hat H_{kl} 
  + \sum_{<kl>:ver} \hat V_{kl} \label{eqbb} 
\end{equation}
This equation holds for any choice of the functions $L_k,R_k,U_k,D_k$. 
Of course the hard part is finding a choice of these functions 
such that (\ref{eqb}) is an $m$-potential. We will find a choice for which
$\hat E_k \ge 8, \hat H_{kl} \ge 0 , \hat V_{kl} \ge 0 $, and 
these lower bounds are all attained by the strip configurations.

Some explanation of the left,right,up,down terminology is in order. 
Each of the functions $L_k,R_k,U_k,D_k$ is a function of the 16 spins in the
4 by 4 square. However, they depend on these 16 spins only through the 
value of $E_k$ and the four spins along one of the four edges of the square. 
Left, right, up and down refer to which edge. Unfortunately this 
causes some confusion when one considers $\hat H_{kl}$ and $\hat V_{kl}$. 
In $\hat H_{kl}$, $k$ is the left square and $l$ is the right square. 
Now $H_{kl}$ depends on the spins along the right edge of $k$ and the 
left edge of $l$. So we subtract $R_k$ and $L_l$  in the definition of 
$\hat H_{kl}$.

We give the definition of $R_k$, along with a bit of motivation.
The definitions of the other three functions are trivially obtained 
by rotation. We denote the four sites on the right edge of square $k$
by 1,2,3,4. (See figure \figd .) 
In the strip states the four spins $(\s_1,\s_2,\s_3,\s_4)$ can be 
$(++++)$, $(----)$,$(+--+)$, or $(-++-)$. 
If $E_k < 8$ and the four spins agree with one of these four 
configurations, then $R_k=0$.
If $E_k < 8$ and the four spins do not agree with any of 
these four cases, then $R_k=1$. 
If $E_k=8$, then $R_k=0$, regardless of the values of 
$\s_1,\s_2,\s_3,\s_4$. 
Note that in a strip configuration, $L_k=R_k=U_k=D_k=0$. 

\ifodd \preprint 
{
\epsfbox[-150 -10 200 140]{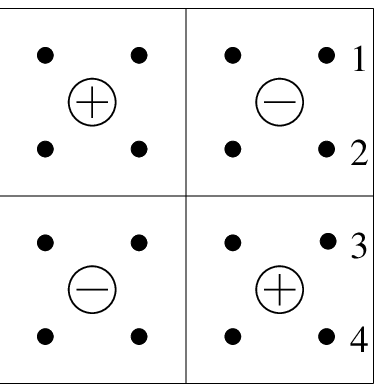}

{\leftskip=20 pt \rightskip= 20 pt \noindent
\no Figure \figd : The definition of $R_k$ depends on the value of $E_k$
and the four spins labelled 1,2,3,4 in the figure. 
\par}

\bigskip
\bigskip
}
\else { } \fi

The motivation for this part of the definition of $R_k$ is to make 
$\hat E_k \ge 8$ for those configurations which have $E_k < 8$. 
As an example, consider figure \figc  (b). This configuration has $E_k=6$.
But $L_k=R_k=U_k=D_k=1$, so $\hat E_k=10$. If we completed the 
definition of $R_k$ by defining it to be zero whenever $E_k >8$,
we would find that $\hat H_{kl}$ can be negative. An example  is 
shown in figure \fige. For the left square $E_k=9$, and for the 
right square $E_l=7$. Obviously $H_{kl}=0$. But $L_l=1$, so if 
$R_k$ were defined to be 0, then $\hat H_{kl}=0$ would be negative. 
To fix this problem we need to make $R_k$ negative in some cases. 
Of course we must do this in such a way that $\hat E_k$ is still bounded 
below by 8. 

\ifodd \preprint 
{
\epsfbox[-90 -10 200 140]{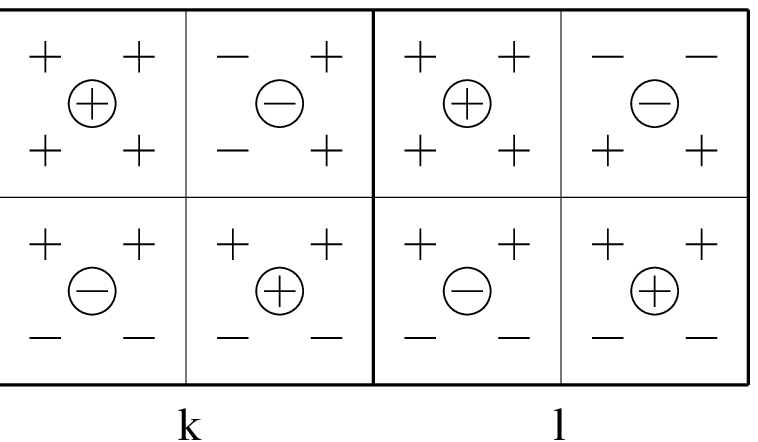}

{\leftskip=20 pt \rightskip= 20 pt \noindent
\no Figure \fige : An example which shows why we must make $R_k$ negative
for some spin configurations.
\par}

\bigskip
\bigskip
}
\else { } \fi

We now complete the definition of $R_k$. If $E_k>8$ we set $R_k= -1$
if three or more of the spins at $1,2,3,4$ are different from the block spin
of the block in which they sit. Otherwise, $R_k=0$. 
Note that $R_k$ is still zero in all four strip states. 
We want to show that 
\begin{equation}
\hat E_k \ge 8, \quad \hat H_{kl} \ge 0 , \quad 
\hat V_{kl} \ge 0 \label{eqc}
\end{equation}
This is easily done on a computer. With the majority rule constraint,
the number of allowable configurations on a 2 by 2 block is 11. 
So for a 4 by 4 block it is $11^4=14,641$. This is small enough that
we can verify $\hat E_k \ge 8 $ by simply computing every case.
The number of cases to be checked for the second and third inequalities
is $(14,641)^2=214,358,881$. Luckily, they do not all need to be checked. 
Since $H_{kl} \ge 0$, we have $\hat H_{kl} \ge 0$ if  $R_k+L_l \le 0$. 
This reduces the number of cases that must be explicitly computed to 
something managable. 
The result of the computer program is that (\ref{eqc}) is indeed true. 
Thus we have succeeded in writing the Hamiltonian as an $m$-potential.

Finally, we need to determine the ground states. The four strip states 
simultaneously minimize each of $\hat E_k$, $\hat H_{kl}$ and 
$\hat V_{kl}$. We will show that these are the only configurations 
that do this. We start by asking what configurations on a square have
$\hat E_k=8$. According to the computer there are 76 such configurations,
including the 4 strip states. We now consider a square $k$ and
the four squares adjacent to it. (We label them as follows:
$i=$up, $j=$right, $l=$down, $m=$ left.)
We ask for what configurations on square $k$ is it possible
to find configurations on square $i,j,l,m$ such that
\begin{eqnarray}
  && \hat E_k=\hat E_i=\hat E_j=\hat E_l=\hat E_m=0 \nonumber \\
  && \hat H_{kj}=\hat H_{mk}=\hat V_{ik} = \hat V_{kl} =0 \label{eqzz}
\end{eqnarray}
Again, we enlist the help of the computer. 
The answer is that there are only four such configurations on square $k$,
the four strip states. Thus in a ground state every 4 by 4 square is
one of the strip states. In the strip states $L_k=R_k=U_k=D_k=0$, so 
(\ref{eqzz}) implies 
$H_{kj}=H_{mk}=V_{ik} =V_{kl} =0$. This implies that in a ground state
we have the same strip state in every 4 by 4 square. 
This completes the proof of proposition 3.
\qed

\medskip

Let $\s^\prime_{cb}$ denote a checkerboard block spin configuration. 
We have shown that when the Hamiltonian $H(\s)$ is restricted to $\s$ 
with $T(\s,\s^\prime_{cb}) \ne 0$, then $H(\s)$ 
is an $m$-potential with four ground states.
It follows from a theorem of Holsztynski and Slawny \cite{hs}
that this restricted $H$ satisfies the sort of Peierls condition that 
one needs to carry out Pirogov-Sinai theory. 
Pirogov-Sinai theory implies that  the original spin system conditioned 
on the checkerboard block spin configuration will have four Gibbs states
at low temperature that are small perturbations of the four ground states.
The conditioned system we must consider is not simply given by restricting
$H(\s)$ to the configurations with $T(\s,\s^\prime_{cb}) \ne 0$.
We must also add $-\ln T$ to the Hamiltonian. The Ising Hamiltonian comes 
with a factor of $\beta$, so $-\ln T$ is a small perturbation which 
can be handled by the Pirogov-Sinai theory.
(Pirogov-Sinai theory is needed here rather than just a simple Peierls 
argument because the rotational symmetry of the lattice is being broken
as well as the global spin flip symmetry.
An introduction to Pirogov-Sinai theory in the context in which we need it
may be found in appendix B of \cite{efs}. The original reference is \cite{ps}.
See also \cite{bi},\cite{kp},\cite{sin},\cite{zah}.

\ifodd \preprint 
{
\epsfbox[0 0 200 280]{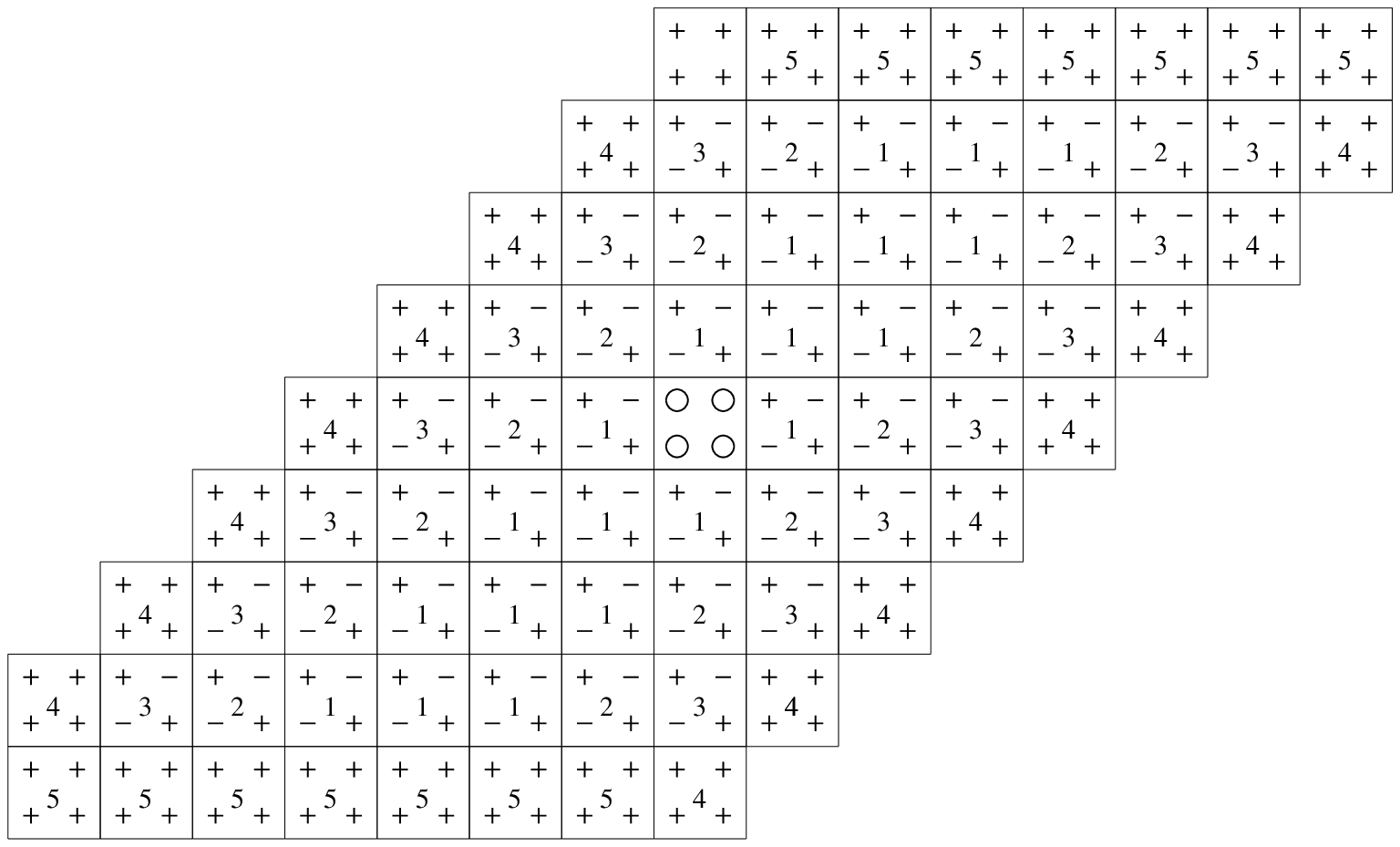}

{\leftskip=20 pt \rightskip= 20 pt \noindent
\no Figure \figz : The special block spin configuration that picks out
one of the pure phases for the original spin system conditioned on the 
checkerboard block spin configuration. Note that only block spins are
shown. We leave it to the reader to draw in the strip state 
that is picked out. The four block spins that
are ``unfixed'' are indicated by circles. 
\par}

\bigskip
\bigskip
}
\else { } \fi

Pirogov-Sinai theory establishes the phase
transition that is responsible for the renormalized measure being
non-Gibbsian. However, we are not finished. To apply the argument of 
\cite{efs}, we must show that one of the four pure phases may be selected by a 
suitable choice of boundary conditions. This would be easy if we were 
free to choose the boundary condition for the original spins. Unfortunately,
we must show that this pure phase can be selected by a choice of 
boundary condition for the block spins. 
The special block spin configuration we use to select a pure phase is 
shown in figure \figz. Note that it is based on a finite volume in the 
shape of a parallelogram rather than a square. The two horizontal boundaries
of the parallelogram favor one of the four strip states.
The two boundaries at a 45 degree angle with respect to the lattice 
directions are neutral in the sense that each of the four strip states
has the same boundary energy where it hits these boundaries.
To prove that this block spin configuration does indeed pick out one
strip state, we show that the Hamiltonian can be written as 
an m-potential with a unique ground state. The strategy is the same as 
before, but we must modify the definition of the functions 
$L_k,R_k,U_k$ and $D_k$ in the vicinity of the boundary. The details 
are provided in the appendix.

Finally, we need to show that by changing the boundary condition for
the block spins, we can change the distribution of the block spins near
the origin. This then implies that the renormalized measure cannot
be quasilocal, i.e., is not consistent with any quasilocal specification.
To do this we consider four block spins arranged in a square located 
near the origin. In the language of \cite{efs} we ``unfix'' these four 
block spins. 
The four block spins we unfix are indicated in figure \figz by circles.
Inside the four blocks the original spins will all be $+1$ 
with probability $1-O(e^{-\beta})$. Let $S$ denote the set containing 
the sites of the four block spins that were unfixed.
Let $\chi$ be the indicator function
for the event that all four of these block spins are 
equal to $+1$. So $\chi=1(\sigma^\prime_i = +1, i \in S)$. 
Let $E^\prime$ denote expectation with respect to the renormalized 
measure $\mu T$ on the block spins. 
$E^\prime(\chi| S^c)$ denotes the conditional
expectation of $\chi$ where we condition on the block spins not in $S$. 
This is a function of the block spins, so we write it as 
$E^\prime(\chi| S^c)(\sigma^\prime)$. 
(Of course, it actually only depends on $\s^\prime_i$ with $i \notin S$.)
Let $U_+$ be the set of block spin configurations which agree with the 
block spin configuration shown in figure \figz and 
are arbitrary outside of the region shown in the figure.
This is an open set in the product topology.
We have shown that 
\begin{equation}
E^\prime(\chi| S^c)(\sigma^\prime) = 1 - O(e^{-\beta}), \quad \forall 
\s^\prime \in U_+
\end{equation}

Now suppose that we modify \figz as follows. We increase the height of the
parallelogram by two block spins by moving each of the horizontal sides of 
the parallelogram by one block spin. The border of the parallelogram 
still consists of plus block spins, and the interior is the same checkerboard
configuration as in figure \figz. But now we will get the
pure phase which is the global spin flip of the pure phase we had before.
Hence the four block spins in $S$ will be $-1$ with probability 
$1-O(e^{-\beta})$. 
So 
\begin{equation}
E^\prime(\chi| S^c)(\sigma^\prime) = O(e^{-\beta}), \quad \forall \s^\prime
\in U_-
\end{equation}
where $U_-$ is the set of block spin configurations 
which agree with this modified version of figure $\figz$,
and are arbitrary outside of the parallelogram.
The above estimates are uniform in the size of the parallelogram, so 
this proves that the conditional distribution of the four block spins
we unfixed is essentially discontinuous and so cannot come from a quasilocal 
specification. 

Our proof that the majority rule transformation
is not defined at low temperature for 2 by 2 
blocks is driven by the phase transition that takes place for the 
system of original spins with the constraint given by the checkboard
block spin configuration. 
Our proof requires that the temperature be very low. However, one might 
expect that the transformation will not be defined for all temperatures
below the critical temperature of this constrained system.
Monte Carlo calculations of Ould-Lemrabott 
indicate that the critical $\beta$ is approximately 1.0 \cite{ol}. 
(For comparison, the critical $\beta$ of the unconstrained 
Ising model is about $0.44$.) 
Cirillo and Olivieri studied a slightly different majority 
rule transformation with 2 by 2 blocks \cite{co}. 
When there is a tie in the block
they take the block spin to equal the spin in the upper left corner of
the block. They found that the critical $\beta$ for the constraint
of the checkerboard block spin configuration is approximately $1.6$.  
Benfatto, Marinari, and Olivieri did a Monte Carlo  study
of a renormalization group
transformation in which the block spin is equal to the sum of the spins
in the block \cite{bmo}. 
For the block spin configuration in which all the block spins are
zero, they found that the constrained system's critical $\beta$
was only about 10\% higher than than of the original unconstrained 
model.

\newpage

\no {\bf 3. Triangular lattice at $T=0$}

\bigskip

In this section we prove theorem 2. For the triangular lattice 
the blocks used by the majority rule are in fact triangles containing 3 sites.
We will continue to refer to these triangles as blocks. 
Given a block spin configuration
the ground state of the original spins subject to the majority rule
constraint imposed by the block spin configuration need not be unique. 
Luckily, this possible degeneracy (which can be rather large) does not
concern us. To compute the minimum in (\ref{eqb}) we need only find one 
ground state. We will give an algorithm for constructing one ground state.
The algorithm will be local - the spin at a site is determined by 
the block spin for the block of that site and the block spins of 
the six blocks that surround that block.

\ifodd \preprint 
{
\epsfbox[-60 50 200 300]{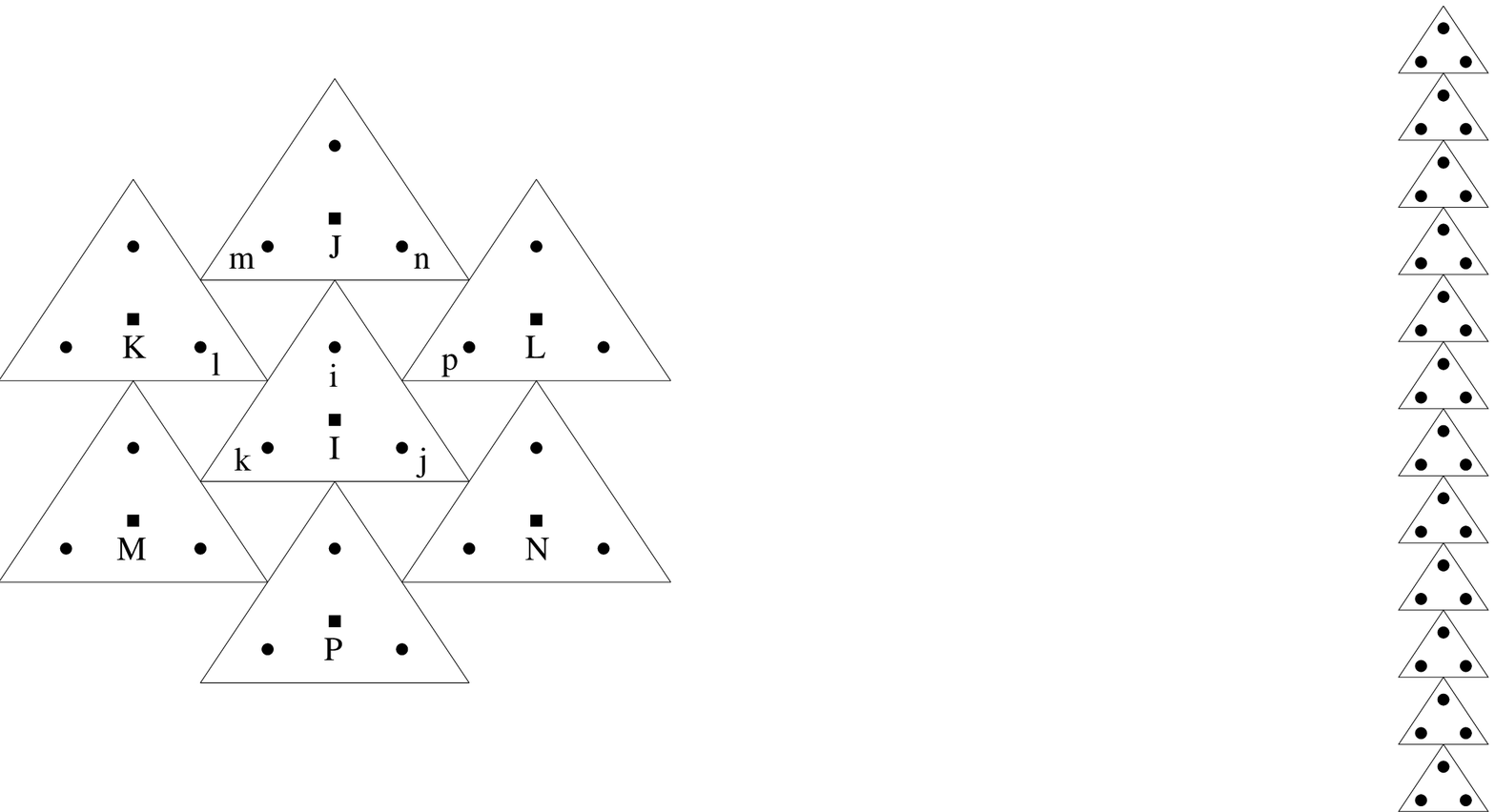}

{\leftskip=20 pt \rightskip= 20 pt \noindent
\no Figure \figf : Labelling of the original spins (circles) and block spins
(squares) used in proposition 4 and table 1.
\par}

\bigskip
\bigskip
}
\else { } \fi

Consider site $i$ in figure \figf. The four block spin sites 
closest to site $i$ are labelled $I,J,K,L$ in the figure.
We need to give these block spins names. We will refer to $I$ as the 
``block spin of site $i$''.
Now suppose we stand at block spin $I$ and face site $i$. 
Block spin $J$ is the closest block spin in the forward direction,  
so we call it the ``forward block spin of site $i$''. 
Block spin $K$ is the closest block spin to the left, so we refer to it
as the ``left block spin of site $i$''. $L$ will be called the 
``right block spin of site $i$''.
We are going to first show that it is possible to find a ground state
with the following property for every site. If at least one of the forward,
left or right block spins agrees with the block spin of the site
then the spin at the site agrees with the block spin of the site.
As an example, consider figure \figf. If $\s^\p_I= +1$ and at least one of
$\s^\p_J,\s^\p_K,\s^\p_L$ equals $+1$, then $\s_i$ is $+1$.
(We are not claiming that every ground state has this property, only
that at least one does.)
In the following proposition 
we break this property into two properties for the sake of the proof.

\medskip

\no {\bf Proposition 4:} There is a ground state with the following properties.

\no (I) If $i$ is a site with its block spin equal to its forward block
spin, then $\s_i$ equals the block spin of $i$. 

\no (II) If $i$ is a site with its block spin equal to either
its left block spin or its right block spin,
then $\s_i$ equals the block spin of $i$.

\medskip

\no {\bf Proof:} We start by showing that there is a ground state with 
property I. It is enough to consider the case $\s^\p_I=\s^\p_J=+1$. 
Suppose $\s_i= -1$. By the majority rule constraint we must have
$\s_j=\s_k= +1$. Since $\s^\p_J= +1$, the majority rule constraint
implies at least one of $\s_m$ and $\s_n$ is $+1$.
Thus at least three of the nearest neighbors of $\s_i$ are $+1$. 
So changing $\s_i$ from $-1$ to $+1$ will not raise the energy. 
Applying this argument where needed, we obtain a ground state with property I. 

Now we take a ground state with property I, and show that we can obtain a 
ground state with property II. 
It suffices to consider the case $\s^\p_I=\s^\p_K= +1$. 
By property I, this implies $\s_l= +1$. 
Suppose $\s_i = -1$. Then $\s_j=\s_k= +1$. 
Thus at least three of the nearest neighbors of $\s_i$ are $+1$. 
So changing $\s_i$ from $+1$ to $-1$ will not raise the energy. 
\qed 

\medskip

Given a block spin configuration, the above two properties 
determine the ground state at a site unless the block spin of that 
site is different from all three of the forward, left and right block
spins of the site.
Consider figure \figf and define three conditions:

\no (Ci) $\s^\p_J=\s^\p_K=\s^\p_L= - \s^\p_I$

\no (Cj) $\s^\p_N=\s^\p_L=\s^\p_P= - \s^\p_I$

\no (Ck) $\s^\p_M=\s^\p_K=\s^\p_P= - \s^\p_I$

\no If (Ci) does not hold, then the two properties in proposition 4
determine $\s_i$.
Likewise, they determine $\s_j$ unless (Cj) holds, and  
determine $\s_k$ unless (Ck) holds. 
Suppose that (Ci) holds and to be concrete consider the case 
$\s^\p_J=\s^\p_K=\s^\p_L= -1$ and $\s^\p_I= +1$.
In this case, property I implies that $\s_m=\s_n= -1$. Property II 
implies $\s_l=\s_p = -1$. Thus at least four of the nearest 
neighbors of $\s_i$ are $-1$. So if $\s_i= +1$, we can lower the energy
by changing it to $-1$. However, if one of $\s_j$ or $\s_k$ is $-1$, then
the majority rule constraint does not permit such a change. 
If neither of (Cj) or (Ck) hold, then $\s_j=\s_k=+1$, and so the 
ground state must have $\s_i= -1$. 
If two or more of (Ci), (Cj) or (Ck) hold, then for one of the sites for which
the condition holds the corresponding spin must be -1.
We showed above that when condition (Cx) holds, the four nearest neighbors
of site $x$ outside of the block containing $x$ are all opposite to the 
block spin of $x$. So when two of these conditions hold, we will have 
the same energy no matter which site we choose to put the $-1$ at. 

We now have an explicit algorithm for finding a ground state. 
Given the block spins 
$\s^\p_I,\s^\p_J,\s^\p_K,\s^\p_L,\s^\p_M,\s^\p_N,\s^\p_P$,
the spins $\s_i,\s_j,\s_k$ are determined as follows. 

\no 1. If (Cx) does not hold, then set $\s_x=\s^\p_I$, where 
$x=i,j,k$. 

\no 2. If (Cx) holds and the other two of (Ci), (Cj) and (Ck) do not, 
then set $\s_x= -\s^\p_I$.

\no 3. If two or more of  (Ci), (Cj) and (Ck) hold, then set 
$\s_x = - \s^\p_I$ for one of the $x$ for which (Cx) holds and 
set $\s_y= \s^\p_I$ for the other two sites.
(This step is ambigious, but we can remove the ambiguity by making
some arbitrary rule for the choice of the site $x$.)

The above algorithm is local. To determine the value of an original 
spin at a site, we only need to know the values of the block spins in a
neighborhood of that site. Thus the ground state energy is a local function
of the block spins. We compute it as follows. Consider the 12 blocks
shown in figure \figg. Given the values the block spins for these 
12 blocks, our algorithm determines the values of the original spins
which are in the three inner blocks.
The 9 nearest neighbor bonds
shown in the figure are chosen so that when they are translated by all 
translations commensurate with the block spin lattice, 
we get every bond in the original lattice exactly once. Thus the 
ground state energy $H^\p(\s^\p)$ 
is obtained by computing the energy of these 9 bonds 
and then summing over translations. 
Obviously the support of a term in $H^\p$ must be a subset of the block
spins shown in figure \figg, or a translation of this set.
In fact, we find that the only terms that actually appear in $H^\p$ 
are those with support contained in some set of 7 block spins 
arranged in a hexagon along with the block at the center of the hexagon.
For example, the seven block spins shown in figure \figf are such
a hexagonal set.

\ifodd \preprint 
{
\epsfbox[-100 80 230 320]{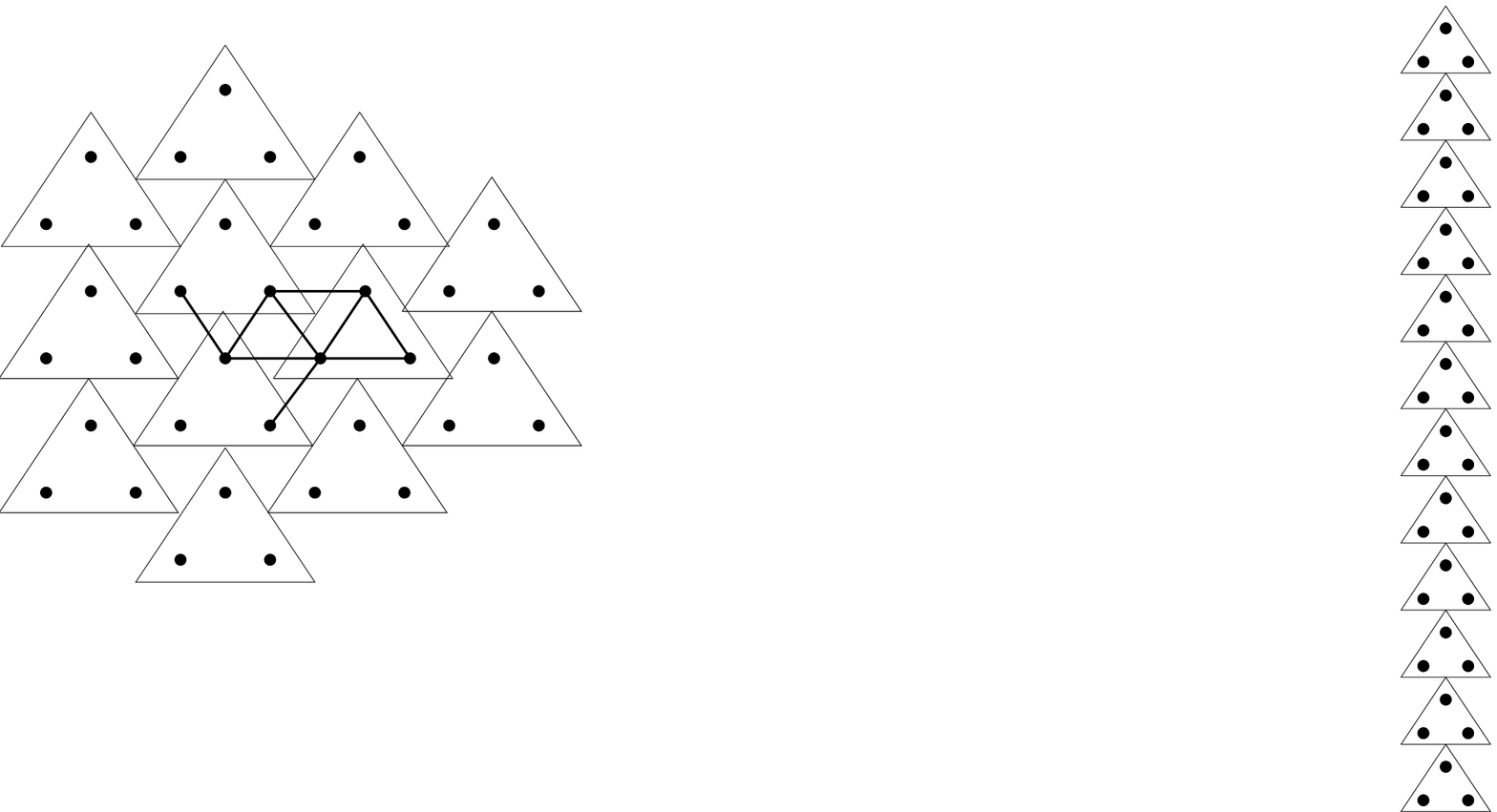}

{\leftskip=20 pt \rightskip= 20 pt \noindent
\no Figure \figg : Picture used in computing the local Hamiltonian
$H^\p$. 
\par}

\bigskip
\bigskip
}
\else { } \fi

While the original triangular lattice is invariant under rotations by 
a multiple of 60 degrees, the blocking partially breaks this symmetry 
and so $H^\p$ need only be invariant under rotations by multiples of 
120 degrees. In addition to this rotational symmetry, the blocked lattice
is also invariant under some reflections. Taking these symmetries
into account, $H^\p$ has 15 different terms. 
They are given in table 1. Only one element from each symmetry class is
given in the table. 

The arguments we have given apply in any finite volume with periodic 
boundary conditions provided the volume is not so small that 
the sorts of regions we have been considering wrap back around on 
themselves. Thus we have proved theorem 2. 

\bigskip

\ifodd \preprint 
{
\begin{tabular}{||l|r||} \hline
 Set of sites  &  Coeffecient   \\ \hline
 I,J         &  26/16    \\
 L,K         &   2/16    \\
 J,P         &  -3/16    \\
 I,J,L,K     &  -4/16    \\
 J,L,P,K     &  -3/16    \\
 J,N,P,M     &   1/16    \\
 I,L,P,K     &   5/16    \\
 I,J,N,M     &  -1/16    \\
 I,J,L,P     &   3/16    \\
 J,L,N,P     &  -1/16    \\
 L,N,M,K     &  -1/16    \\
 I,J,P,M     &   1/16    \\
 J,L,N,P,M,K &   1/16    \\
 I,L,N,P,M,K &   1/16    \\
 I,J,L,N,M,K &  -1/16    \\  \hline
\end{tabular}


\bigskip

{\leftskip=20 pt \rightskip= 20 pt \noindent
\no Table 1: The terms in the local Hamiltonian $H^\p$. The sites 
are labelled as in figure \figf. Only one term from each symmetry class
is shown. 
\par}
}
\else { } \fi

\vskip 1.5 in

\no {\bf Acknowledgements: } This work was supported in part by 
NSF grant DMS-9303051.

\newpage

\no {\bf Appendix }

\bigskip
\bigskip

In this appendix we consider the system of original spins with 
the constraint imposed by the block spin configuration shown in 
figure \figz from section two. We want to show that the Hamiltonian
is an $m$-potential and has a unique ground state in the interior of 
the parallelogram.
The strategy is the same as in section two. The only difference is
in the definition of the functions $L_k,R_k,D_k$ and $U_k$. 
These functions now depend on where the square $k$ is in relation 
to the parallelogram in figure \figz. We divide the squares into
five types, labelled 1 to 5, according to where the square is.
See figure \figz. The definitions of the functions are a bit involved.
We will not attempt to motivate them; they were found mainly with
trial and error and a little intuition. We should emphasize that they
are by no means unique.

To define the functions $L_k,R_k,U_k,D_k$, we introduce a little notation.
Recall that each of these functions is a function of a four by four square
in the original lattice. For function $L_k,R_k,U_k$ or $D_k$,
the ``four boundary spins'' will refer to the 
four spins along the left, right, upper or lower edge of the square, 
respectively. Each original spin belongs to a two by two 
majority rule block. We will refer of the block spin of that two by two 
block as the block spin associated to the original spin. Now let
$n_\boundary$ be the number of the four boundary spins that are not 
equal to their associated block spin. Let $m_\boundary$ be the number 
of the four boundary spins that are equal to $-1$. 
Finally we define a variable $strip$ that takes on the values true and
false. $strip$ is true if the four boundary spins agree with 
the four boundary spins in some strip configuration, i.e., they 
must be one of $(++++)$, $(----)$,$(+--+)$, or $(-++-)$. 

If square $k$ is of type $1$, then the functions
$L_k,R_k,U_k,D_k$ are all defined as follows. If $E_k<8$ and $strip$ is 
not true, then the value is $+1$. If $E_k>8$ and $n_\boundary>2$, then the
value is $-1$. Otherwise the value is $0$. 

Now consider a square $k$ is of type $2$ near the left boundary. 
The functions $R_k$ and $D_k$ are defined as they were for type
$1$ squares.
The functions $L_k$ and $U_k$ are defined as follows.
If $E_k<8$ and $n_\boundary=0$, then the value is $+1$. 
If $E_k>8$ and $n_\boundary>2$ then the
value is $-1$. Otherwise the value is $0$. 
The definitions for a type $2$ square near the right boundary 
are obtained in the obvious way by symmetry considerations.

For a square $k$ of type $3$ located near the left boundary the definitions
are as follows.
$L_k$ and $U_k$ equal $m_\boundary$ 
when $m_\boundary \le 3$ and equal $3$ when $m_\boundary=4$.
The definition of $R_k$ and $D_k$ is a bit more complicated when $k$ 
is of type $3$. If $E_k+L_k+U_k \le 9$ and $strip$ is not
true, then their value is $+1$. 
If $E_k \ge 10$ and $n_\boundary>2$, then their value is $-1$. 
The definitions for a type $3$ square near the right boundary 
are obtained in the obvious way by symmetry considerations.

If square $k$ is of type $4$ and near the left boundary, 
then $R_k$ and $D_k$ equal $-n_\boundary$ 
when $n_\boundary \le 3$ and 
equal $-3$ when $n_\boundary=4$.
The functions $L_k,U_k$ are identically zero when $k$ is of type $4$.  
Again, symmetry determines the definitions when the square is near
the right boundary.

If square $k$ is of type $5$ and near the top boundary, 
then $D_k$ equals $-1$ when $E_k>0$ and 
equals $0$ otherwise.
The functions $L_k,R_k,U_k$ are identically zero.
If square $k$ is of type $5$ and near the bottom boundary, 
then $U_k$ equals $-1$ when $E_k>0$ and 
equals $0$ otherwise.
The functions $L_k,R_k,D_k$ are identically zero.

In the strip state selected by the block spin configuration in 
figure \figz, we have $E_k=8,8,10,0,0$
for $k$ of type $1,2,3,4,5$, respectively. We also have 
$H_{kl}=V_{kl}=0$ and $L_k=R_k=U_k=D_k=0$ in this strip state except 
for the following cases. 
If $k$ is type 4 and $l$ is type $3$ and
$k$ is immediately left of $l$, then $H_{kl}=2$ and $L_l=2$. 
Thus in the strip state shown, $\hat E_k=8,8,8,0,0$
for $k$ of type $1,2,3,4,5$, respectively, and $\hat H_{kl}=\hat V_{kl}=0$. 
To prove that our decomposition of the Hamiltonian is an $m$-potential
we must show that these values are in fact the minimum of each of 
these functions. As in section two, this is easily done on the computer.

Finally we ask what are the ground states of the Hamiltonian.
By the results of section two, 
in the interior of the parallelogram the configuration
must be in one of the four strip states. 
Now consider the top horizontal edge of the parallelogram.
For $l$ of type 5 we have $E_l=0$, so the original spins in the 4 by 4 
squares of type 5 must all be $+1$. Now let $k$ be a type 1 square just 
below a type 1 square $l$. 
Then $E_k=8$ and so $U_k=0$. 
We also have $D_l=0$, so $\hat V_{kl}=V_{kl}$. But $\hat V_{kl}=0$, 
so $V_{kl}=0$. Since the spins in square $l$ are all $+1$, it follows 
that a particular strip state is picked out for square $k$.
The same argument applies to the bottom edge of the parallelogram.
The height of the parallelogram is chosen so that the strip state
picked out by the top edge is the same as the one picked out by the 
bottom edge.
We have not shown that in a ground state the configuration 
must look like figure \figz along the edges of the parallelogram
at 45 degrees to the lattice directions. In fact, it need not. 
There are ways to modify figure \figz near these two edges that do not raise
the energy. What we have shown is that any such modification cannot
lower the energy, and away from these two edges the ground state must
be the strip state picked out by the horizontal edges of the 
parallelogram.

\ifodd \preprint 
{}
\else 
{ 
\newpage

\centerline {Table}

\begin{tabular}{||l|l|r||} \hline
 Set of sites  &  Coeffecient &   Symmetry factor  \\ \hline
 I,J         &  26/16   &  3 \\
 L,K         &   2/16   &  3 \\
 J,P         &  -3/16   &  3 \\
 I,J,L,K     &  -4/16   &  3 \\
 J,L,P,K     &  -3/16   &  3 \\
 J,N,P,M     &   1/16   &  3 \\
 I,L,P,K     &   5/16   &  1 \\
 I,J,N,M     &  -1/16   &  1 \\
 I,J,L,P     &   3/16   &  6 \\
 J,L,N,P     &  -1/16   &  6 \\
 L,N,M,K     &  -1/16   &  3 \\
 I,J,P,M     &   1/16   &  6 \\
 J,L,N,P,M,K &   1/16   &  1 \\
 I,L,N,P,M,K &   1/16   &  3 \\
 I,J,L,N,M,K &  -1/16   &  3 \\  \hline
\end{tabular}

\bigskip

{\leftskip=20 pt \rightskip= 20 pt \noindent
\no Table 1: The terms in the local Hamiltonian $H^\p$. The sites 
are labelled as in figure \figf. Only one term from each symmetry class
is shown. 
\par}

\newpage

\centerline {Figure Captions}

\bigskip
\bigskip

\no \figa. Circled spins are block spins, uncircled spins are original
spins. The block spins are in the checkerboard configuration, and the 
original spins are in one of the four ground states which we call
``strip states''.

\smallskip

\no \figb. The original Hamiltonian is the sum of the three
types of terms $E_k$, $H_{kl}$, and $V_{kl}$. The block spins are not shown
here.

\smallskip

\no \figc. Example showing that the strip states do not 
give the minimum of $E_k$. 
The configuration in (b) has lower $E_k$ than the strip state shown
in (a). 

\smallskip

\no \figd. The definition of $R_k$ depends on the value of $E_k$
and the four spins labelled 1,2,3,4 in the figure. 

\smallskip

\no \fige. An example which shows why we must make $R_k$ negative
for some spin configurations.

\smallskip

\no \figz. The special block spin configuration that picks out
one of the pure phases for the original spin system conditioned on the 
checkerboard block spin configuration. Note that only block spins are
shown. We leave it to the reader to draw in the strip state 
that is picked out. The four block spins that
are ``unfixed'' are indicated by circles. 

\smallskip

\no \figf. Labelling of the original spins (circles) and block spins
(squares) used in proposition 4 and table 1.

\smallskip

\no \figg. Picture used in computing the local Hamiltonian
$H^\p$. 

\newpage
\epsfbox{mrlowt_figa.eps}

\newpage
\epsfbox{mrlowt_figb.eps}

\newpage
\epsfbox{mrlowt_figc.eps}

\newpage
\epsfbox{mrlowt_figd.eps}

\newpage
\epsfbox{mrlowt_fige.eps}

\newpage
\epsfbox{mrlowt_figz.eps}

\newpage
\epsfbox{mrlowt_figf.eps}

\newpage
\epsfbox{mrlowt_figg.eps}

} 
\fi

\newpage

\newcommand \jsp {{\it J. Stat. Phys.} }
\newcommand \cmp {{\it Commun. Math. Phys.}}
\newcommand \prl {{\it Phys. Rev. Lett.}}

\end{document}